\documentstyle[12pt,epsfig]{article}
\def\be{\begin{equation}} \def\ee{\end{equation}} \def\bea{\begin{eqnarray}}
\def\eea{\end{eqnarray}} \def\nnb{\nonumber}

\begin{document}

\hfill{October 18, 2023,\ \ \ {\tt rdO16\_ver2}}

\begin{center}
\vskip 6mm 
\noindent
{\Large\bf  
Radiative decay of the sub-threshold $1_1^-$ and $2_1^+$ states of $^{16}$O
in cluster effective field theory
}
\vskip 6mm 

\noindent
{\large 
Shung-Ichi Ando\footnote{mailto:sando@sunmoon.ac.kr}, 
}
\vskip 6mm
\noindent
{\it
Department of Display and Semiconductor Technology, 
Sunmoon University,
Asan, Chungnam 31460,
Republic of Korea
}
\end{center}

\vskip 6mm

Radiative decay of the sub-threshold $1_1^-$ and $2_1^+$ states of $^{16}$O
is studied in cluster effective field theory.
The wave function normalization factors
for initial and final states of the radiative decay amplitudes
are deduced
by using the phase shift data 
of elastic $\alpha$-$^{12}$C scattering 
for $l=0,1,2$, which are related to 
the asymptotic normalization coefficients of $0_1^+$, $1_1^-$, $2_1^+$ 
states of $^{16}$O for the two-body $\alpha$-$^{12}$C channel;
then only one unfixed parameter remains 
in each of the radiative decay amplitudes.
We fit the parameters to the experimental radiative decay rates and  
apply the fitted parameters to the study of
radiative $\alpha$ capture on $^{12}$C, 
$^{12}$C($\alpha$,$\gamma$)$^{16}$O.
The order of magnitude of astrophysical $S$ factors of 
$E1$ and $E2$ transitions of 
$^{12}$C($\alpha$,$\gamma$)$^{16}$O
is reproduced compared to the experimental data, 
and we discuss an improvement in the calculation 
of $S$ factors of $^{12}$C($\alpha$,$\gamma$)$^{16}$O. 

\vskip 5mm
\noindent PACS(s):
11.10.Ef, 
24.10.-i, 
25.55.-e, 
26.20.Fj  

\newpage 
\vskip 2mm \noindent
{\bf 1. Introduction}

The radiative $\alpha$ capture on $^{12}$C, $^{12}$C($\alpha$,$\gamma$)$^{16}$O,
is a fundamental reaction in nuclear astrophysics, which determines, 
together with the triple $\alpha$ process,
the C/O ratio in the core of a helium-burning star~\cite{f-rmp84}. 
It plays a crucial role 
in simulations of nucleosynthesis in star 
evolutions~\cite{ww-pr93,ietal-aj01},
e.g., 
the radiative $\alpha$ capture reaction is 
one of the most uncertain nuclear processes 
in different models
of type Ia Supernova nucleosynthesis~\cite{jj-20}. 
The $\alpha$ capture rate, or alternatively the astrophysical $S$ 
factor of $^{12}$C($\alpha$,$\gamma$)$^{16}$O at the energy
of helium-burning process, namely the Gamow-peak energy, $E_G=0.3$~MeV, has 
never been measured in an experimental facility because of the Coulomb barrier. 
One needs to employ a theoretical model, whose parameters are fitted to 
experimental data measured at a few MeV energy, and extrapolate the 
reaction rate down to $E_G$.
During the last half-century, many experimental and theoretical 
studies have been carried out; refer to, e.g., 
Refs.~\cite{bb-npa06,cetal-epja15,bk-ppnp16,detal-rmp17,a-epja21} for review. 

During the last decade, we have been studying an estimate of the $S$
factor of $^{12}$C($\alpha$,$\gamma$)$^{16}$O by employing the methodology
of effective field theory (EFT)~\cite{w-physica79,bvk-arnps02,dgh}. 
An EFT is constructed by introducing a scale that separates relevant 
degrees of freedom at low energy from irrelevant degrees of freedom at 
high energy. Then, one constructs the most general form of effective 
Lagrangian so as to satisfy the symmetry requirement and expands it 
in powers of the number of derivatives; the theory provides us with 
a perturbative expansion scheme in terms of $Q$/$\Lambda_H$ where  
$Q$ is a typical momentum scale of a reaction in question and $\Lambda_H$
is a momentum scale of high energy for separation. The irrelevant 
degrees of freedom are integrated out of the Lagrangian and their effect
is embedded in the coefficients of terms of Lagrangian. 
Those coefficients, 
in principle, can be determined within its mother theory while they,
in practice, are fixed by using empirical values or fitted to experimental
data. Furthermore, the inclusion of a resonance state in theory was studied
by, e.g., Galman~\cite{g-prc09} 
and Habashi, Fleming, and van Kolck~\cite{hfvk-epja21}.
EFTs are applied to various studies for nuclear reactions at low energies,
such as radiative neutron capture on a proton 
for big-bang nucleosynthesis~\cite{cs-prc99,r-npa00,aetal-prc06},
proton-proton capture in hydrogen burning~\cite{kr-npa99,bc-plb01,aetal-plb08}, 
and solar neutrino reactions on a deuteron~\cite{bck-prc01,ash-prc20}. 

To construct an EFT for the calculation of the astrophysical $S$ factor of 
$^{12}$C($\alpha$,$\gamma$)$^{16}$O at $E_G$, 
we choose the energy difference between the 
open channels of $\alpha$-$^{12}$C and $p$-$^{15}$N states,
$\Delta E=4$~MeV, for the separation scale, 
and one has the large momentum scale $\Lambda_H$ 
as $\Lambda_H = \sqrt{2\mu \Delta E} = 160$~MeV where
$\mu$ is the reduced mass of $\alpha$ and $^{12}$C.  
The relevant degrees of freedom for the theory are $\alpha$ and $^{12}$C 
represented as non-relativistic point-like scalar fields. 
The effective Lagrangian is constructed so as to be invariant under the 
Galilean and gauge transformations, and expanded in terms
of the number of covariant derivatives. 
The typical momentum scale at $E_G$ is $Q=\sqrt{2\mu E_G}=40$~MeV, and the 
expansion parameter becomes $Q/\Lambda_H = 1/4$. 
We introduce auxiliary fields for bound and resonant 
states of $^{16}$O to carry out a momentum expansion around 
the unitary limit~\cite{k-npb97,bs-npa01,ah-prc05},
which reproduces the expression of effective range expansion~\cite{b-pr49}.
In addition, it is straightforward to introduce the electromagnetic
and weak interactions in the theory. 
In the previous works, we constructed an EFT for an estimate of $S$ factor
of $^{12}$C($\alpha$,$\gamma$)$^{16}$O at $E_G$ studying 
elastic $\alpha$-$^{12}$C scattering for various cases of including 
the bound and resonant states 
of $^{16}$O~\cite{a-epja16,a-prc18,a-jkps18,a-prc23}, 
$E1$ transition of $^{12}$C($\alpha$,$\gamma$)$^{16}$O~\cite{a-prc19}, 
and $\beta$ delayed $\alpha$ emission from $^{16}$N~\cite{a-epja21}.  

In this work, we study the radiative decay of excited $1_1^-$ and $2_1^+$ 
states of $^{16}$O within the cluster EFT. 
The main aim of the present work is to seek constraints on the 
radiative capture amplitudes of $^{12}$C($\alpha$,$\gamma$)$^{16}$O. 
Feynman diagrams of the radiative decay process are parts of 
those of $^{12}$C($\alpha$,$\gamma$)$^{16}$O, and one can fix 
the coupling constants of $O^*\gamma O$ vertex functions
by using the experimental data of decay rates
of $1_1^-$ and $2_1^+$ states of $^{16}$O. 
We also employ the wave function normalization factors, equivalently
the asymptotic normalization coefficients (ANCs) of 
$0_1^+$, $1_1^-$, $2_1^+$ states of $^{16}$O, to fix the coupling constants
of $\alpha CO$ vertex function for the final $0_1^+$ ground state of $^{16}$O
and $O^*\alpha C$ vertex functions for the initial $1_1^-$ and $2_1^+$ states
of $^{16}$O. 
Moreover, we improve the regularization method for 
logarithmic divergence appearing from loop diagrams.
In the previous work~\cite{a-prc19}, we employed two regularization methods,
sharp cutoff regularization and dimensional regularization.
In this work, we apply the dimensional regularization method to 
the calculation of logarithmic divergence, which previously was 
regularized by introducing a sharp cutoff. 
Then, we fit the couplings of $O^*\gamma O$ vertex functions to 
the experimental data of decay rates of $1_1^-$ and $2_1^+$ states of 
$^{16}$O as functions of the sharp cutoff parameter $r_C$; a mild
cutoff dependence is found compared to the result in the previous 
study~\cite{a-prc19}.
Now there are no free parameters in the amplitudes of 
$^{12}$C($\alpha$,$\gamma$)$^{16}$O. 
We calculate the $S$ factors of 
$E1$ and $E2$ transitions of $^{12}$C($\alpha$,$\gamma$)$^{16}$O by
using the fitted parameters and 
plot the curves of $S$ factors with the experimental data. We find that
the order of magnitude of $S$ factors is reproduced and discuss how
the calculation of $S$ factors of 
$^{12}$C($\alpha$,$\gamma$)$^{16}$O
can be improved.  

The present work is organized as follows. 
In Sec. 2 the expression of effective Lagrangian is reviewed, and 
in Sec. 3, the radiative decay amplitudes are derived considering the 
regularization methods of loop diagrams and 
the wave function normalization factors, 
equivalently the ANCs of $0_1^+$, $1_1^-$,
$2_1^+$ states of $^{16}$O. In Sec. 4, the numerical results of fitting the 
parameters and plotting the $S$ factors of 
$^{12}$C($\alpha$,$\gamma$)$^{16}$O are discussed, and finally in 
Sec. 5, the results and discussion of this work are presented.
In Appendix A, masses and formulae of the decay rates for 
the present study are summarized,
and in Appendix B, the derivation of ANC of ground $0_1^+$ state 
of $^{16}$O from the phase shift of elastic $\alpha$-$^{12}$C scattering 
is discussed. 

\vskip 2mm \noindent 
{\bf 2. Effective Lagrangian}

As discussed in the introduction, the theory has a perturbative scheme in 
terms of the number of derivatives: the expansion parameter is obtained as 
$Q/\Lambda_H = 1/4$. While we discussed a modification of the counting rules
for the effective range expansion for elastic $\alpha$-$^{12}$C scattering 
at low energies~\cite{a-prc18}; one or two orders of magnitude larger 
contributions (than the term estimated by using a phase shift datum 
at the lowest energy in the experiment~\cite{tetal-prc09}\footnote{
The lowest energy of the experimental data for the elastic $\alpha$-$^{12}$C 
scattering is $E_\alpha = 2.6$~MeV where $E_\alpha$ is the $\alpha$ energy 
in the lab frame. 
}) 
appear from the Coulomb self-energy term.
To subtract those unnaturally large contributions, 
we include the effective range terms up to $p^6$ order as counter terms.
In the study of $E1$ transition of $^{12}$C($\alpha$,$\gamma$)$^{16}$O, 
on the other hand, it is known that the $E1$ transition is suppressed 
between initial and final isospin singlet states. 
The suppression factor can be seen in the term, 
$(Z_\alpha/m_\alpha - Z_C/m_C)$, in the amplitudes,
where $Z_\alpha$ and $m_\alpha$ 
($Z_C$ and $m_C$) are the proton number and mass of $\alpha$ ($^{12}$C);
if one ignores the binding energies of nuclei, $m_\alpha \simeq 4 m_N$ and 
$m_C \simeq 12 m_N$ where $m_N$ is the nucleon mass, and $Z_\alpha = 2$ 
and $Z_C=8$, the term $(Z_\alpha/m_\alpha - Z_C/m_C)$ vanishes. 
The prime contribution to the $E1$ transition amplitudes comes out of
the coupling constant of $O^*\gamma O$ vertex function. This constant is 
supposed to contain dynamics from high energy: a significant contribution 
may come out of the isovector channel, namely, $p$-$^{15}$N state. 
The $p$-$^{15}$N open channel can appear in loop diagrams instead
of $\alpha$-$^{12}$C state. While the $p$-$^{15}$N state is integrated out
and its effect is embedded in the coupling constant of $O^*\gamma O$ vertex
function; we fitted the coupling constant to experimental data
of $S$ factor for $E1$ transition of $^{12}$C($\alpha$,$\gamma$)$^{16}$O
in the previous study~\cite{a-prc19}. 

The effective Lagrangian for radiative decay of $1_1^-$ and $2_1^+$ states
of $^{16}$O is the same as that for $E1$ and $E2$ transitions of 
$^{12}$C($\alpha$,$\gamma$)$^{16}$O~\cite{a-prc19}. 
Thus, one has
\bea
{\cal L} &=& \phi_\alpha^\dagger \left(
iD_0
+\frac{\vec{D}^2}{2m_\alpha}
+ \cdots
\right) \phi_\alpha
+ \phi_C^\dagger\left(
iD_0
+ \frac{\vec{D}^2}{2m_C}
+\cdots
\right)\phi_C
\nnb \\ && +
\sum_{n=0}^3
C_{n}^{(0)}d^{\dagger}
\left[
iD_0
+ \frac{\vec{D}^2}{2(m_\alpha+m_C)}
\right]^n d^{}
- y_{}^{(0)}\left[
d_{}^{\dagger}(\phi_\alpha 
\phi_C)
+ (\phi_\alpha 
\phi_C)^\dagger d^{}_{}
\right] 
\nnb \\ && +
\sum_{n=0}^3
C_{n}^{(1)}d_{i}^\dagger
\left[
iD_0
+ \frac{\vec{D}^2}{2(m_\alpha+m_C)}
\right]^n d_{i}
- y_{}^{(1)}\left[
d_{i}^\dagger(\phi_\alpha O_i^{(1)} \phi_C)
+ (\phi_\alpha O_i^{(1)} \phi_C)^\dagger d_{i}
\right] 
\nnb \\ && +
\sum_{n=0}^3
C_{n}^{(2)}d_{ij}^\dagger
\left[
iD_0
+ \frac{\vec{D}^2}{2(m_\alpha+m_C)}
\right]^n d_{ij}
- y_{}^{(2)}\left[
d_{ij}^\dagger(\phi_\alpha O_{ij}^{(2)} \phi_C)
+ (\phi_\alpha O_{ij}^{(2)} \phi_C)^\dagger d_{ij}
\right] 
\nnb \\ &&
- h^{(1)}\frac{y^{(0)}y^{(1)}}{\mu}\left[
(
{\cal O}_l^{(1)} d^{}
)^\dagger
d_l + \mbox{\rm H.c.}
\right]
-h^{(2)}\frac{y^{(0)}y^{(2)}}{\mu^2}\left[
({\cal O}_{ij}^{(2)} d^{} 
)^\dagger
d_{ij} + \mbox{\rm H.c.}
\right]
+ \cdots
\,,
\eea
where $\phi_\alpha$ ($m_\alpha$) and
$\phi_C$ ($m_C$) are fields (masses) of $\alpha$ and $^{12}$C,
respectively.
$D_\mu$ is the covariant derivative,
$D_\mu \phi = (\partial_\mu + ieZA_\mu)\phi$, where $e$ is the electric charge,
$Z$ is the number of protons in a nucleus, and $A_\mu$ is the photon field. 
The combination of the terms, $iD_0 + \vec{D}^2/(2m)$,
is required for the invariance under the Galilean transformation.
The dots denote higher-order terms.
$d$, $d_i$, $d_{ij}$ are fields of composite states of $\alpha$ and $^{12}$C
for $l=0,1,2$, respectively, which represent the bound 
states of $^{16}$O~\footnote{
We suppress the terms of Lagrangian for resonance states of $^{16}$O
for the sake of simplicity, which are involved when fitting the phase
shift data of elastic $\alpha$-$^{12}$C scattering. 
One can find the expression of the terms in Eq.~(6) 
in Ref.~\cite{a-prc23}.
}, 
where the spin states are represented as Cartesian tensors of lank $l$.
The coefficients $C_n^{(l)}$ correspond with the 
effective range parameters in the form of $C_n^{(l)}/y^{(l)2}$ with $l=0,1,2$,
where $y^{(l)}$ are the coupling constants of $O^*\alpha C$ 
and $O\alpha C$ vertex functions. 
As mentioned before, we include the effective range parameters 
up to $p^6$ order;
namely, $n=0,1,2,3$.  
$h^{(1)}$ and $h^{(2)}$ are bare coupling constants of contact 
$O^*\gamma O$ vertex functions, with which the infinities from loop diagrams
are renormalized. 
The projection operators in $p$-wave and $d$-wave read 
\bea
&&
O_{l}^{(1)} =
i\frac{\stackrel{\leftrightarrow}{D}_i}{M}
\equiv
i \left(
\frac{\stackrel{\rightarrow}{D}_C}{m_C} -
\frac{\stackrel{\leftarrow}{D}_\alpha}{m_\alpha} 
\right)_i
\,,
\ \ \
O_{ij}^{(2)} =
-\frac{\stackrel{\leftrightarrow}{D}_i}{M}
\frac{\stackrel{\leftrightarrow}{D}_j}{M}
+ \frac13\delta_{ij}
\frac{\stackrel{\leftrightarrow}{D}^2}{M^2}
\,,
\\
&& {\cal O}_i^{(1)} = \frac{iD_i}{m_O}\,,
\ \ \
{\cal O}_{ij}^{(2)} = -\frac{1}{m_O^2}(D_iD_j - \frac{1}{3}\delta_{ij}D^2)\,,
\eea
where $m_O$ is the mass of ground $0_1^+$ state of $^{16}$O. 

\vskip 2mm \noindent
{\bf 3. Radiative decay amplitudes}

In this section, we derive expressions of the decay amplitudes from 
the effective Lagrangian.
We discuss an improvement in regularization methods of the log divergence
from loop diagrams and the calculation of wave function normalization 
factors by using the phase
shift data of elastic $\alpha$-$^{12}$C scattering; 
final expressions of the decay amplitudes of $1_1^-$ and $2_1^+$ states
of $^{16}$O contain only one parameter for each of the amplitudes,
and the parameters are fitted to the experimental decay rates in the 
next section. 

\vskip 2mm \noindent
{\bf 3.1 Amplitudes}

\begin{figure}
\begin{center}
  \includegraphics[width=13cm]{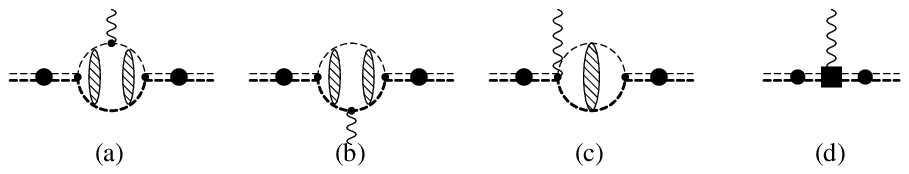}
\caption{
Diagrams of radiative decay amplitudes for 
$^{16}$O$^*$($1_1^-$) $\to ^{16}$O$_{g.s.}+\gamma$
and $^{16}$O$^*$($2_1^+$) $\to ^{16}$O$_{g.s.}+\gamma$.
A thick (thin) dashed line represents a propagator of $^{12}$C ($\alpha$),
and a thick and thin double-dashed line with a filled circle
represents the initial and final states of $^{16}$O, and 
a wavy line the outgoing photon. 
A shaded blob represents
a set of diagrams consisting of all possible one-potential-photon-exchange
diagrams up to infinite order and no potential-photon-exchange one,
namely, the Coulomb greens function. 
A filled box represents the contact interactions 
of $O^*\gamma O$ vertex functions and renormalize the divergence terms
from the loop diagrams.  
}
\label{fig;amplitudes}       
\end{center}
\end{figure}

The radiative decay amplitudes $A^{(l)}$ with $l=1,2$ 
are calculated from diagrams 
displayed in Fig.~\ref{fig;amplitudes}, which are parts of the diagrams
for $^{12}$C($\alpha$,$\gamma$)$^{16}$O
(displayed in Fig.~2 in Ref.~\cite{a-prc19}). 
The amplitudes $A^{(l)}$ with $l=1,2$ are represented as 
\bea
A^{(l=1)} &=& 
\vec{\epsilon}_{(\gamma)}^*\cdot \vec{\epsilon}_{(l=1)} Y^{(l=1)}\,,
\\
A^{(l=2)} &=& 
\epsilon_{(\gamma)}^{*i}\hat{k}_2^j \epsilon_{(l=2)}^{ij} Y^{(l=2)}\,,
\eea
with 
\bea
Y^{(l)} &=& Y_{(a+b)}^{(l)} + Y_{(c)}^{(l)} + Y_{(d)}^{(l)}\,,
\eea
where $\vec{\epsilon}_{(\gamma)}^*$ is the polarization vector 
of out-going photon, $\vec{\epsilon}_{(l=1)}$ is the polarization
vector of initial $p$-wave state of $^{16}$O$^*$($1_1^-$),
and $\hat{k}_2$ is a unit vector of the momentum $\vec{k}_2$ 
of out-going photon from the decay of $2_1^+$ state of $^{16}$O, and 
$\epsilon_{(l=2)}^{ij}$ is the symmetric traceless tensor of 
initial $d$-wave state of $^{16}$O$^*$($2_1^+$). 
The amplitudes, $Y_{(a+b)}^{(l)}$, $Y_{(c)}^{(l)}$, $Y_{(d)}^{(l)}$, 
correspond to the diagrams displayed in Fig.~\ref{fig;amplitudes}. 
Masses and decay rates for the present study are summarized 
in Appendix A. 
Thus, the spin-averaged decay rates in Eq.~(\ref{eq;Gamma_l}) become
\bea
\Gamma^{(l=1)} &=& \frac{\alpha_E k_1}{3\pi}|Y^{(l=1)}|^2\,,
\\ 
\Gamma^{(l=2)} &=& \frac{\alpha_E k_2}{5\pi}|Y^{(l=2)}|^2\,,
\eea
where $\alpha_E$ is the fine structure constant, 
and $k_1$ and $k_2$ are the magnitude of photon momenta in Eqs.~(\ref{eq;k1})
and (\ref{eq;k2}) for radiative decay of $1_1^-$ and $2_1^+$ states
of $^{16}$O, respectively. 
The amplitudes, $Y_{(a+b)}^{(l)}$, $Y_{(c)}^{(l)}$, and $Y_{(d)}^{(l)}$, 
are obtained as  
\bea
Y_{(a+b)}^{(l=1)} &=& 
+\frac{1}{9\pi}y^{(0)}y^{(1)}\gamma_1 
\Gamma(1+\kappa/\gamma_0) \Gamma(2+\kappa/\gamma_1)
\nnb \\ && \times 
\int^\infty_{0}drr W_{-\kappa/\gamma_0,\frac12}(2\gamma_0r)
\left[
\frac{\mu Z_\alpha}{m_\alpha} j_0\left(
\frac{\mu}{m_\alpha} k_1r\right)
-\frac{\mu Z_C}{m_C}j_0\left(
\frac{\mu}{m_C}k_1r
\right)
\right]
\nnb \\ && \times \left\{
\frac{\partial}{\partial r}\left[
\frac{W_{-\kappa/\gamma_1,\frac32}(2\gamma_1r)}{r}
\right]
+2\frac{W_{-\kappa/\gamma_1,\frac32}(2\gamma_1r)}{r^2}
\right\}\,,
\label{eq;Y1ab}
\\
Y_{(c)}^{(l=1)} &=&  
\frac{1}{2\pi} y^{(0)}y^{(1)} \mu \left(
\frac{Z_\alpha}{m_\alpha} - \frac{Z_C}{m_C}
\right)\left[
\frac{2\pi}{\mu}J_0^{div}
-2\kappa H(-i\kappa/\gamma_0)
\right]\,,
\label{eq;Y1c}
\\
Y_{(d)}^{(l=1)} &=& 
- h^{(1)} y^{(0)}y^{(1)} \frac{Z_O}{\mu m_O}\,,
\\
Y_{(a+b)}^{(l=2)} &=& 
- \frac{2}{75\pi}
y^{(0)}y^{(2)}
\frac{\gamma_2^2}{\mu} 
\Gamma(1+\kappa/\gamma_0) \Gamma(3+\kappa/\gamma_2)
\nnb \\ && \times 
\int_0^\infty drrW_{-\kappa/\gamma_0,\frac12}(2\gamma_0 r) 
\left[
\frac{\mu Z_\alpha}{m_\alpha}j_1\left(
\frac{\mu}{m_\alpha}k_2r
\right)
+ \frac{\mu Z_C}{m_C} j_1\left(
\frac{\mu}{m_C}k_2r
\right)
\right]
\nnb \\ && \times
\left\{ 
\frac{\partial}{\partial r}\left[
\frac{W_{-\kappa/\gamma_2,\frac52}(2\gamma_2r)}{r}
+ 3 \frac{W_{-\kappa/\gamma_2,\frac52}(2\gamma_2r)}{r^2}
\right]
\right\}\,,
\label{eq;Y2ab}
\\
Y_{(c)}^{(l=2)} &=&  0\,,
\\
Y_{(d)}^{(l=2)} &=& 
+ h^{(2)} y^{(0)}y^{(2)} \frac{Z_O}{\mu^2m_O^2}k_2 \,, 
\label{eq;Y2d}
\eea
where $\Gamma(x)$, $j_l(x)$ ($l=0,1$), and $W_{\alpha,\beta}(x)$ are  
the gamma function, the spherical Bessel function, and the Whittaker function, 
respectively. 
$\mu$ is the reduced mass of $\alpha$ and $^{12}$C, 
$\mu = m_\alpha m_C/(m_\alpha + m_C)$ = 2795~MeV.  
$\gamma_0$, $\gamma_1$, and $\gamma_2$ are the binding momenta 
for $0_1^+$, $1_1^-$, and $2_1^+$ states of $^{16}$O, respectively;
$\gamma_l = \sqrt{2\mu B_l}$ with $l=0,1,2$ where $B_l$ are the binding 
energies. Thus, one has $\gamma_0 = 200.1$~MeV, $\gamma_1 = 15.88$~MeV, 
and $\gamma_2 = 37.00$~MeV.   
$\kappa$ is the inverse of Bohr radius, $\kappa = \alpha_E Z_\alpha Z_C\mu$,
where $Z_\alpha$ and $Z_C$ (and $Z_O$ in Eq.~(\ref{eq;Y2d})) 
are the number of protons in $\alpha$ and $^{12}$C (and $^{16}$O); 
$Z_\alpha=2$ and $Z_C=6$ (and $Z_O=8$), respectively; 
thus, $\kappa = 245$~MeV.
In addition, the function $H(\eta)$ in Eq.~(\ref{eq;Y1c}) reads
\bea
H(\eta) &=& \psi(i\eta) + \frac{1}{2i\eta} -\ln(i\eta)\,,
\eea
where $\psi(z)$ is the digamma function and $\eta$ is the Sommerfeld 
parameter.
(We will mention the function $J_0^{div}$ in Eq.~(\ref{eq;Y1c}) in the
next subsection.)

One may notice that there are five constants, \{$y^{(0)}$, $y^{(1)}$,
$y^{(2)}$, $h^{(1)}$, $h^{(2)}$\}, appearing in the amplitudes.   
Three of them, $y^{(0)}$, $y^{(1)}$, $y^{(2)}$, are determined by
using the wave function normalization factors, and the 
remaining two parameters, $h^{(1)}$ and $h^{(2)}$, are fitted to 
the radiative decay data. Before fixing the parameters, we discuss the 
renormalization of infinities from one-loop diagrams in the 
next subsection. 

\vskip 2mm \noindent
{\bf 3.2 Renormalization of the loop diagrams}

The one-loop diagrams (a), (b), and (c) in Fig.~\ref{fig;amplitudes}
diverge. The divergent terms are subtracted by the counter terms,
$h^{(1)}$ and $h^{(2)}$, in the diagram (d) in Fig.~\ref{fig;amplitudes}.
For the diagrams (a) and (b), the $r$-space integrals in Eqs.~(\ref{eq;Y1ab})
and (\ref{eq;Y2ab}) diverge: 
the log divergence appears in the limit where  $r$ goes to zero. 
In the previous work in Ref.~\cite{a-prc19}, 
we introduced a sharp-cutoff $r_C$ in the $r$-space integral for the 
radiative capture amplitude of $E1$ transition of 
$^{12}$C($\alpha$,$\gamma$)$^{16}$O, 
and the divergent term was renormalized by the counter term, 
$h^{(1)}$.  
We found that the cutoff dependence was severe, and moreover,
it was inconsistent with the regularization method, 
namely the dimensional regularization,
which was applied to the calculation of diagram (c). 
In this work, we separate the $r$-space integrals in Eqs.~(\ref{eq;Y1ab})
and (\ref{eq;Y2ab}) into two parts 
by introducing a cutoff parameter $r_C$, and 
the calculation of the log-diverging integral in the short-range part 
is carried out by employing
the dimensional regularization in $d=4-2\epsilon$ space-time dimensions,
ignoring the other finite terms. 
Thus, a minor cutoff dependence remains in the numerical result. 
The log divergence of the short-range part is replaced as  
\bea
\int_0^{r_C}\frac{dr}{r} &\to& 
\left(
\frac{\mu_{DR}}{2}
\right)^{2\epsilon} 
\int_0^{r_C}dr r^{-1+2\epsilon} = \frac{1}{2\epsilon}
+ \ln\left(
\frac{\mu_{DR}}{2}r_C
\right) + O(\epsilon)\,,
\label{eq;oovr_integral}
\eea
where $\mu_{DR}$ is a scale parameter from the dimensional regularization. 
For the diagram (c) in Fig.~\ref{fig;amplitudes}, 
we employ dimensional regularization and calculate the loop integral
in $d=4-2\epsilon$ space-time dimension; one has the divergence
term $J_0^{div}$ as
\bea
J_0^{div} &=& \frac{\kappa\mu}{2\pi}\left[
\frac{1}{\epsilon} -3C_E + 2 + \ln\left(
\frac{\pi\mu_{DR}^2}{4\kappa^2}
\right)
\right]\,,
\eea
where $C_E$ is the Euler-Mascheroni constant, $C_E=0.5772\cdots$.

We subtract the divergent terms being proportional to $1/(2\epsilon)$ 
as well as some constant terms by using the coupling constants, 
$h^{(1)}$ and $h^{(2)}$, as 
\bea
\lefteqn{
h^{(1)} 
+ \frac{\kappa\mu^2}{9\pi} \frac{m_O}{Z_O}\left(
\frac{Z_\alpha}{m_\alpha} - \frac{Z_C}{m_C}
\right)\left[ 
\left(
\frac{\mu_{DR}}{2}
\right)^{2\epsilon}
\int_0^{r_C}\frac{dr}{r^{1-2\epsilon}}
-  \frac{9\pi}{\kappa\mu} J_0^{div} 
\right]
}
\nnb \\ && = 
h_R^{(1)} 
+ \frac{\kappa\mu^2}{9\pi} \frac{m_O}{Z_O}\left(
\frac{Z_\alpha}{m_\alpha} - \frac{Z_C}{m_C}
\right) \left[
\ln\left(
\frac{\mu_{DR}}{2}r_C
\right)
- 9  \ln\left(
\frac{\mu_{DR}}{2\kappa}
\right)
\right]
\,,
\\
\lefteqn{
h^{(2)} + \frac{2\kappa\mu^3}{75\pi}\frac{m_O^2}{Z_O}
\left(
\frac{Z_\alpha}{m_\alpha^2} + \frac{Z_C}{m_C^2}
\right)
\left(
\frac{\mu_{DR}}{2}
\right)^{2\epsilon}
\int_0^{r_C}\frac{dr}{r^{1-2\epsilon}}
}
\nnb \\ && = 
h_R^{(2)} + \frac{2\kappa\mu^3}{75\pi}\frac{m_O^2}{Z_O}
\left(
\frac{Z_\alpha}{m_\alpha^2} + \frac{Z_C}{m_C^2}
\right)
\ln\left(
\frac{\mu_{DR}}{2}r_C
\right)
\,,
\eea
where $h_R^{(1)}$ and $h_R^{(2)}$ are 
the renormalized coupling constants
and fitted to the experimental radiative decay rates
in the next section. 

\vskip 2mm \noindent
{\bf 3.3 Wave function normalization factors}

\begin{figure}
\begin{center}
  \includegraphics[width=12cm]{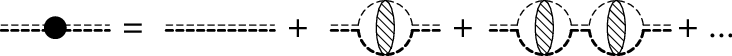}
\caption{
Diagrams for dressed $^{16}$O propagator.
A thick and thin double dashed line with and without a filled circle
represents a dressed and bare $^{16}$O propagator, respectively.
	See the caption of Fig.~\ref{fig;amplitudes} as well. 
}
\label{fig;propagator}       
\end{center}
\end{figure}

The coefficients, \{$y^{(0)}$, $y^{(1)}$, $y^{(2)}$\}, appeared 
in the amplitudes 
can be fixed by using the wave function normalization factors, 
$\sqrt{{\cal Z}_l}$ with $l=0,1,2$,
which are related to the ANCs of $0_1^+$, $1_1^-$, $2_1^+$ states of $^{16}$O
for the two-body $\alpha$-$^{12}$C system.
The relations between $y^{(l)}$ and $\sqrt{{\cal Z}_l}$ may be given 
as~\cite{g-npa04}\footnote{
The wave function normalization factors, $\sqrt{{\cal Z}_l}$ with
$l=0,1,2$, appear only in the initial and final states of reaction 
amplitudes. $y^{(l)}$ appearing in the intermediate states are 
represented, 
excluding $\sqrt{{\cal Z}_l}$, 
as $y^{(l)}=\sqrt{2(2l+1)\pi\mu^{2l-1}}$.  
} 
\bea
y^{(l)} &=& \sqrt{
2(2l+1)\pi\mu^{2l-1}
}\sqrt{
{\cal Z}_l
}\,,
\label{eq;yl}
\eea
with $l=0,1,2$. 
${\cal Z}_l$ are defined  
in the the inverse of dressed $^{16}$O propagators, $D_l(p)$, 
for the two-body $\alpha$-$^{12}$C system, 
where $p$ is the relative 
momentum between $\alpha$ and $^{12}$C.
Diagrams of dressed $^{16}$O propagators are depicted in 
Fig.~\ref{fig;propagator}, 
and one has
\bea
\frac{1}{D_l(p)} &=& 
\frac{1}{K_l(p) -2\kappa H_l(p)} = \frac{{\cal Z}_l}{E + B_l} + \cdots\,, 
\label{eq;invDl}
\eea
where $E$ is the kinetic energy of $\alpha$ and $^{12}$C, 
$E = p^2/(2\mu)$, and $B_l$ are the binding energies of bound states
of $^{16}$O as the $\alpha$-$^{12}$C two-body system. 
The dots denote finite terms at $E=-B_l$. 
The term, $-2\kappa H_l(p)$, is the Coulomb self-energy term 
from the one-loop diagram in Fig.~\ref{fig;propagator}, 
and one has 
\bea
H_l(p) &=& W_l(p) H(\eta)\,,
\\
W_l(p) &=& \left(
\frac{\kappa^2}{l^2} + p^2
\right) W_{l-1}(p)\,, \ \  \ W_0(p) = 1\,.
\eea
The term, $K_l(p)$, is represented in terms of the effective range
parameters as  
\bea
K_l(p) &=& -\frac{1}{a_l} + \frac12 r_l p^2 - \frac14 P_l p^4 
+ Q_l p^6\,,
\eea
where one or two parameters of the effective range parameters are fixed 
by using the condition that 
the inverse of the propagator $D_l(p)$ vanishes at the binding momenta 
$\gamma_l=\sqrt{2\mu B_l}$,  
$D_l(i\gamma_l) = 0$.
Thus, two parameters are fixed 
by using the binding energies of $0_1^+$ 
and $0_2^+$ states of $^{16}$O for $l=0$, and  
one parameter is fixed by using 
the binding energies of $1_1^-$ and $2_1^+$ 
states of $^{16}$O for $l=1,2$.  
We fix 
the scattering length and effective range $a_0$ 
and $r_0$ for $l=0$ and 
the scattering lengths 
$a_l$ for $l=1,2$.
Thus, one has the inverse of propagators as~\cite{a-prc18,a-jkps18} 
\bea
D_0(p) &=&  - \frac14\left[
\gamma_{01}^2\gamma_{02}^2 
+ (\gamma_{01}^2 + \gamma_{02}^2) p^2
+ p^4
\right] P_0
\nnb \\ && 
+ \left[
-\gamma_{01}^4\gamma_{02}^2 
-\gamma_{01}^2\gamma_{02}^4
- (\gamma_{01}^4 + \gamma_{01}^2\gamma_{02}^2 + \gamma_{02}^4)p^2 
+ p^6 
\right] Q_0
\nnb \\ &&
-2\kappa\left[
\frac{\gamma_{02}^2 + p^2}{\gamma_{01}^2 - \gamma_{02}^2} H_0(i\gamma_{01})
-\frac{\gamma_{01}^2 + p^2}{\gamma_{01}^2 - \gamma_{02}^2} H_0(i\gamma_{02})
+ H_0(p)
\right] \,,
\label{eq;D0}
\\
D_l(p) &=& \frac12r_l (\gamma_l^2+p^2) 
+ \frac14P_l(\gamma_l^4-p^4)
+ Q_l (\gamma_l^6+p^6)
+ 2\kappa\left[H_l(i\gamma_l) - H_l(p)
\right]\,,
\label{eq;Dl}
\eea
with $l=1,2$: $\gamma_{01}(=\gamma_0)$ and $\gamma_{02}$ 
are the bounding momenta 
of $0_1^+$ and $0_2^+$ states of $^{16}$O, respectively; 
$\gamma_{02}=\sqrt{2\mu B_{02}}$,
where $B_{02}$ is the binding energy of the $0_2^+$ state of $^{16}$O  as
the two-body $\alpha$-$^{12}$C system,
and one has $\gamma_{02}=78.86$~MeV. 

The other effective range parameters in Eqs.~(\ref{eq;D0}) and 
(\ref{eq;Dl}) are fitted to the phase shift data of elastic 
$\alpha$-$^{12}$C scattering.  
\begin{figure}[t]
\begin{center}
\resizebox{0.2\textwidth}{!}{
 \includegraphics{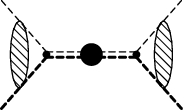}
}
\caption{
Diagram of the scattering amplitude.
Shaded blobs in the initial and final states represent non-perturbative 
Coulomb interaction, namely, the Coulomb wave functions. 
See the captions of Figs.~\ref{fig;amplitudes} and \ref{fig;propagator} 
as well.
}
\label{fig;scattering_amplitude}
\end{center}
\end{figure}
The elastic scattering amplitudes are calculated from a diagram depicted 
in Fig.~\ref{fig;scattering_amplitude}. 
Expression of the $S$ matrices, amplitudes, and fitted values 
of some parameters can be found in Ref.~\cite{a-prc23}. 
Thus, one may calculate the wave function normalization factors 
$\sqrt{{\cal Z}_l}$
from the inverse of propagators $D_l(p)$ in Eq.~(\ref{eq;invDl}) as 
\bea
\sqrt{{\cal Z}_l} &=& \left(
2\mu \left| \frac{\partial D_l(p)}{\partial p^2}
\right|_{p^2=-\gamma_l^2}
\right)^{-1/2} \,,
\eea
and through the relations in Eq.~(\ref{eq;yl}), those factors are multiplied 
to the amplitudes, 
as the coupling constants, $y^{(l)}$, 
for the initial $1_1^-$ or $2_1^+$ state and final $0_1^+$
state of $^{16}$O. 
In addition, the wave function normalization factors are
related to ANCs through the 
formula derived by Iwinski, Rosenberg, and Spruch~\cite{irs-prc84}:
\bea
|C_b|_l &=& 
\frac{\gamma_l^l}{l!} \Gamma(l+1+\kappa/\gamma_l) 
\sqrt{2\mu {\cal Z}_l}\,.
\label{eq;ANCs}
\eea

Now, we rewrite the amplitudes of radiative decay reactions 
in terms of the wave function renormalization factors as
\bea
\lefteqn{Y_{(a+b)}^{(l=1)} = + 
\frac{16}{3\sqrt3}\sqrt{{\cal Z}_{01}{\cal Z}_1}\, 
\Gamma(1+\kappa/\gamma_0)\Gamma(2+\kappa/\gamma_1)\gamma_1^2}
\nnb \\ && \times \int^\infty_{r_C} dr e^{-(\gamma_0+\gamma_1)r}
\gamma_0\gamma_1r^2 U(1+\kappa/\gamma_0,2,2\gamma_0r)
\left[
\frac{\mu Z_\alpha}{m_\alpha}j_0\left(
\frac{\mu}{m_\alpha}k_1r
\right)
- \frac{\mu Z_C}{m_C} j_0\left(
\frac{\mu}{m_C}k_1r
\right)
\right]
\nnb
\label{eq;Yab_1} 
\\ && \times \left[
(3-\gamma_1r)U(2+\kappa/\gamma_1,4,2\gamma_1r)
-2\gamma_1r(2+\kappa/\gamma_1)U(3+\kappa/\gamma_1,5,2\gamma_1r)
\right]\,,
\\
\lefteqn{Y_{(c)}^{(l=1)} = +
\sqrt3\sqrt{{\cal Z}_{01}{\cal Z}_1}\, 
\mu \left(
\frac{Z_\alpha}{m_\alpha} - \frac{Z_C}{m_C}
\right) [-2\kappa H(-i\kappa/\gamma_0)]\,,}
\\
\lefteqn{Y_{(d)}^{(l=1)} = 
- 2\sqrt3\,\pi\sqrt{{\cal Z}_{01}{\cal Z}_1}\frac{Z_O}{\mu m_O}
}
\nnb \\ && \times
\left\{
h_R^{(1)} 
+ \frac{\kappa\mu^2}{9\pi} 
\frac{m_O}{Z_O}\left(
\frac{Z_\alpha}{m_\alpha} - \frac{Z_C}{m_C}
\right)
\left[
\ln\left(
\frac{\mu_{DR}}{2}r_C
\right)
- 9 \ln\left(
\frac{\mu_{DR}}{2\kappa}
\right)
\right]
\right\}\,,
\\
\lefteqn{Y_{(a+b)}^{(l=2)} = 
- \frac{64\sqrt5}{75}\sqrt{{\cal Z}_{01}{\cal Z}_2} \,
\gamma_2^3\, \Gamma(1+\kappa/\gamma)
\Gamma(3+\kappa/\gamma_2) 
}
\nnb \\ &&
\times \int_{r_C}^\infty dr e^{-(\gamma_0 + \gamma_2)r} 
\gamma_0 \gamma_2^2r^3 U(1+\kappa/\gamma_0,2,2\gamma_0 r)
\left[
\frac{\mu Z_\alpha}{m_\alpha} j_1\left(
\frac{\mu}{m_\alpha}k_2r
\right)
+ \frac{\mu Z_C}{m_C}j_1\left(
\frac{\mu}{m_C}k_2r
\right)
\right]
\nnb \\ && \times
\left[
(5-\gamma_2r)U(3+\kappa/\gamma_2,6,2\gamma_2 r) 
- 2\gamma_2 r (3+\kappa/\gamma_2)U(4+\kappa/\gamma_2,7,2\gamma_2 r)
\right]\,,
\label{eq;Yab_2}
\\
\lefteqn{Y_{(c)}^{(l=2)} = 0\,,
}
\\
\lefteqn{Y_{(d)}^{l=2)} = + 2\sqrt5\,\pi\sqrt{{\cal Z}_{01}{\cal Z}_2} 
\frac{Z_Ok_2}{\mu m_O^2}
\left[
h_R^{(2)}
+ \frac{2\kappa\mu^3}{75\pi}
\frac{m_O^2}{Z_O}\left(
\frac{Z_\alpha}{m_\alpha^2} + \frac{Z_C}{m_C^2}
\right)\ln\left(
\frac{\mu_{DR}}{2}r_C
\right)
\right] \,,
}
\eea
where we have also rewritten the Whittaker function, $W_{\kappa,\mu}(z)$, 
in terms of the Kummer function, $U(a,b,c)$. 
One may notice that the combinations for the ANCs appear 
in the amplitudes, $Y_{(a+b)}^{(l=1)}$ 
and $Y_{(a+b)}^{(l=2)}$, in Eqs.~(\ref{eq;Yab_1}) and (\ref{eq;Yab_2}). 

\vskip 2mm \noindent
{\bf 4. Numerical results}

In this section, we first briefly review the calculation of 
the wave function renormalization 
factors, 
$\sqrt{{\cal Z}_{01}}$, 
$\sqrt{{\cal Z}_{1}}$, 
$\sqrt{{\cal Z}_{2}}$, 
by fitting the effective range parameters to
the experimental phase shift data of elastic $\alpha$-$^{12}$C 
scattering at low energies.
In the following, we quote the values of the ANCs 
rather than those of $\sqrt{{\cal Z}_l}$
because the ANCs are more reliable quantities 
for the other theoretical studies. 
Then, we fit the remaining two parameters, $h_R^{(1)}$ and $h_R^{(2)}$, 
to the radiative decay rates of sub-threshold $1_1^-$ and $2_1^+$ states 
of $^{16}$O.
After fitting the parameters, there is no free parameter in 
$E1$ and $E2$ transitions of amplitudes of $^{12}$C($\alpha$,$\gamma$)$^{16}$O,
and we perform a parameter-free calculation of the astrophysical $S$ factors 
for $E1$ and $E2$ transitions
of $^{12}$C($\alpha$,$\gamma$)$^{16}$O.
We will see that the order of magnitude of the $S$ factors 
compared to the experimental data is reproduced.

As discussed in the previous section, the coupling constants, \{$y^{(0)}$,
$y^{(1)}$, $y^{(2)}$\}, are fixed by using the wave function normalization 
factors, $\sqrt{{\cal Z}_{01}}$, $\sqrt{{\cal Z}_1}$, $\sqrt{{\cal Z}_2}$,
which are related to the ANCs through the relation in Eq.~(\ref{eq;ANCs}).
In the previous work, we obtained the ANCs for $1_1^-$ and 
$2_1^+$ states of $^{16}$O as~\cite{a-prc23}
\bea
|C_b|_1 &=& 1.727(3) \times 10^{14}\ \ \textrm{fm}^{-1/2}\,,
\label{eq;ANC,11m}
\\
|C_b|_2 &=& 3.1(6) \times 10^{4}\ \ \textrm{fm}^{-1/2}\,,
\label{eq;ANC,21p}
\eea   
by fitting the effective range parameters to the phase shift 
data~\cite{tetal-prc09}. 
We employ the fitted values of effective range parameters in TABLE II 
in Ref.~\cite{a-prc23} to calculate $\sqrt{{\cal Z}_1}$ 
and $\sqrt{{\cal Z}_2}$.   
The ANC of $1_1^-$ state agrees with the other results
in literature while that of $2_1^+$ state is about four times smaller than
the other ones; the fit of effective range parameters for $l=2$ is  
sensitive to the conditions of effective range
parameters imposed on the low-energy region where the experimental 
phase shift data are not available~\cite{a-prc22}. 

To calculate $\sqrt{{\cal Z}_{01}}$, we need to fix the values of two
effective range parameters, $P_0$ and $Q_0$, in Eq.~(\ref{eq;D0});
we refit the parameters to the phase shift data including the 
$0_1^+$, $0_2^+$,
$0_3^+$, $0_4^+$ states of $^{16}$O. Details of the fit of
the parameters $P_0$ and $Q_0$ are discussed in Appendix B. 
Thus, 
we have the ANC of ground $0_1^+$ state of $^{16}$O as
\bea
|C_b|_{01} = 44.5(3) \ \ \textrm{fm}^{-1/2}\,,
\label{eq;ANC,01p}
\eea
where the corresponding value of $y^{(0)}$ is
$y^{(0)} = 0.355(3)$~MeV$^{-1/2}$.
Those quantities, $|C_b|_{01}$ and $y^{(0)}$, are related through
Eqs.~(\ref{eq;yl}) and (\ref{eq;ANCs});
$|C_b|_{01} = \frac{\mu}{\sqrt{\pi}} \Gamma(1+\kappa/\gamma_{01}) y^{(0)}$.
It may be worth pointing out that this value of $|C_b|_{01}$ is close 
to that used in the recent $R$-matrix analysis for 
$^{12}$C($\alpha$,$\gamma$)$^{16}$O reported by deBoer et al.,
$|C_b|_{01}= 58$~fm$^{-1/2}$~\cite{detal-rmp17}.
The reported values of the ANC, $|C_b|_{01}$, in literature
are still scattered, 13.9(24) to 4000~fm$^{-1/2}$; 
see TABLE XVI in Ref.~\cite{detal-rmp17}. 
In addition, we fitted the values of $y^{(0)}$, along with the constant 
$h_R^{(1)}$ to the experimental data of $S_{E1}$ factor 
of $^{12}$C($\alpha$,$\gamma$)$^{16}$O 
by using the sharp cutoff regularization 
scheme in the previous work~\cite{a-prc19} and had  
$y_0=0.253(9)$ -- $2.249(84)$~MeV$^{-1/2}$ with the cutoff values 
$r_C=0.01$ -- $0.35$~fm. 
\begin{table}
\begin{center}
\begin{tabular}{ l | l | l l }
\hline 
$r_C$~(fm) & $h_R^{(1)}\times 10^{-5}$~(MeV$^3$) &  
 $h_R^{(1)}$~(MeV$^3$)\cite{a-prc19} & $y^{(0)}$~(MeV$^{-1/2}$)\cite{a-prc19} 
 \cr \hline 
0.005 & 2.503(2), 2.635(2) &  -- & -- \cr
0.01  & 2.626(2), 2.757(2) & 5.268(1)$\times 10^5$ & 0.253(9) \cr
0.035 & 2.775(2), 2.906(2) & 2.448(1)$\times 10^5$ & 0.310(11) \cr
0.05  & 2.825(2), 2.957(2) & 1.529(1)$\times 10^5$ & 0.347(12) \cr
0.10  & 2.957(2), 3.088(2) & -0.070(1)$\times 10^5$ & 0.495(18) \cr \hline
\end{tabular}
\caption{ 
Two values of $h_R^{(1)}$ in the second column are obtained 
by fitting to the radiative decay rate of $1_1^-$ state of $^{16}$O, 
$\Gamma_1^{exp}$, as a function of the cutoff parameter $r_C$ 
employing the dimensional regularization 
where the scale parameter $\mu_{DR}$ is fixed as $\mu_{DR} = \Lambda_H$. 
Values of $h_R^{(1)}$ and $y^{(0)}$ in the third and fourth columns 
are fitted values in the previous work~\cite{a-prc19}, where
$h_R^{(1)}$ and $y^{(0)}$ are treated as free parameters and 
fitted to the experimental data of $S_{E1}$ factor of radiative $\alpha$
capture on $^{12}$C employing the sharp cutoff regularization method. 
}
\label{table;h1R}
\end{center}
\end{table}
(The result of fitted values of $h_R^{(1)}$ and $y^{(0)}$ 
for the cutoff values $r_C = 0.01 - 0.10$~fm 
is included in Table~\ref{table;h1R}.) 
The $y^{(0)}$ values are converted to the ANC as
$|C_b|_{01} = 31.7(1) - 282(10)~\textrm{fm}^{-1/2}$.
One may see that the value of ANC in Eq.~(\ref{eq;ANC,01p}) is reproduced
at $r_C \simeq 0.05$~fm.  
In this work, we fix the value of $y^{(0)}$ by using the ANC of ground
$0_1^+$ state of $^{16}$O in Eq.~(\ref{eq;ANC,01p})
in the radiative decay amplitudes.   

Now, the coupling constants $h_R^{(1)}$ and $h_R^{(2)}$ are fitted to  
the experimental data of radiative decay rates of 
$1_1^-$ and $2_1^+$ states of $^{16}$O~\cite{twc-npa93},
\bea
\Gamma^{exp}_1 &=& 0.055(3)\ \ \textrm{eV}\,,
\\
\Gamma^{exp}_2 &=& 0.097(3)\ \ \textrm{eV}\,,
\eea
as a function of the cutoff $r_C$ where the scale parameter $\mu_{DR}$ 
is fixed as $\mu_{DR}=\Lambda_H$. 
The short-range part of integrals in the amplitudes, 
$Y_{(a+b)}^{l=1)}$ and $Y_{(a+b)}^{(l=2)}$ in Eqs.~(\ref{eq;Yab_1}) 
and (\ref{eq;Yab_2}), respectively, are replaced by the $1/r$ integrals
in Eq.~(\ref{eq;oovr_integral}); 
the finite components are ignored and
the cutoff dependence remains in the numerical results. 
The fitted values of $h_R^{(1)}$ and $h_R^{(2)}$ are
presented in Tables \ref{table;h1R} and \ref{table;h2R}, respectively.     
The fitted values of $h_R^{(1)}$ and $y^{(0)}$ in the 
previous work~\cite{a-prc19} are displayed in Table \ref{table;h1R} as well. 
\begin{table}
\begin{center}
\begin{tabular}{ l | l }
\hline 
$r_C$~(fm) & $h_R^{(2)}\times 10^{-11}$~(MeV$^4$) \cr \hline 
0.005 & 8.414(2), 8.627(2) \cr
0.01  & 7.759(2), 7.971(2) \cr
0.035 & 7.013(2), 7.226(2) \cr
0.05  & 6.774(2), 6.986(2) \cr
0.10  & 6.146(2), 6.359(2) \cr \hline
\end{tabular}
\caption{
Values of $h_R^{(2)}$ are obtained by fitting 
to the radiative decay rate of $2_1^+$ 
state of $^{16}$O, $\Gamma_2^{exp}$, 
as a function of the cutoff parameter $r_C$.
See the caption of Table \ref{table;h1R} as well. 
}
\label{table;h2R}
\end{center}
\end{table}
We find that there are two values of $h_R^{(1)}$ and $h_R^{(2)}$ with
small error bars. This indicates a large cancelation between the amplitude
of the loop diagrams and that of the counter term
resulting in two same-size amplitudes with different signs.  
The cutoff dependence remains in the fitted values of $h_R^{(1)}$ 
and $h_R^{(2)}$ while that of $h_R^{(1)}$ becomes 
much milder than the previous fitted values obtained by means of the sharp 
cutoff regularization method. 

\begin{figure}
\begin{center}
  \includegraphics[width=13cm]{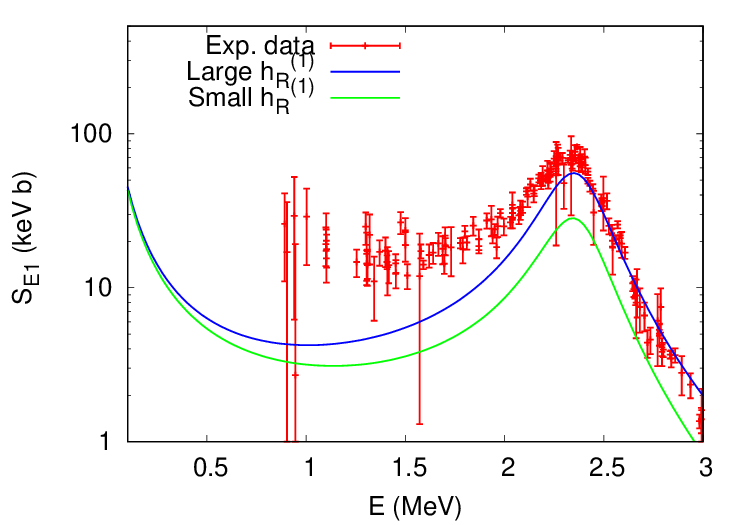}
\caption{
Curves of $S_{E1}$ factor of radiative $\alpha$ capture on $^{12}$C are plotted
as a function of $\alpha$-$^{12}$C energy $E$ in the center-of-mass frame 
by using the large and small values of $h_R^{(1)}$ at $r_C = 0.01$~fm
in Table \ref{table;h1R}. Experimental data are displayed in the figure 
as well.  
}
\label{fig;se1}       
\end{center}
\end{figure}

\begin{figure}
\begin{center}
  \includegraphics[width=13cm]{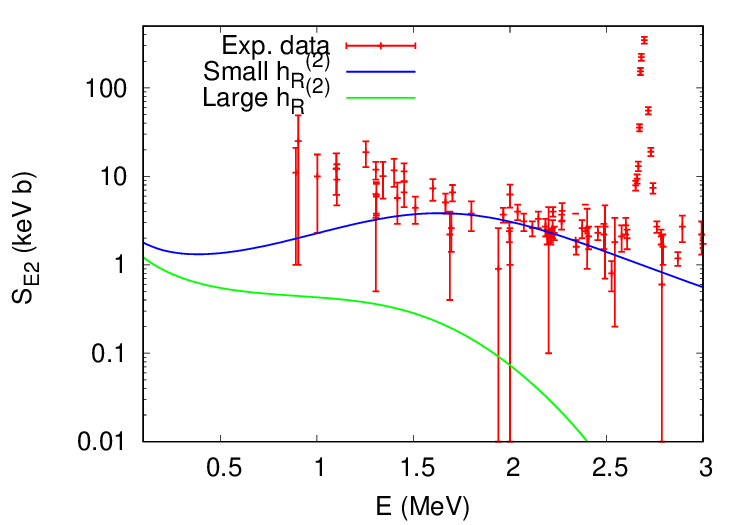}
\caption{
Curves of $S_{E2}$ factor of radiative $\alpha$ capture on $^{12}$C are plotted
as a function of $\alpha$-$^{12}$C energy $E$ in the center-of-mass frame
by using the small and large values of $h_R^{(2)}$ at $r_C = 0.01$~fm
in Table \ref{table;h2R}. Experimental data are displayed in the figure
as well.
}
\label{fig;se2}       
\end{center}
\end{figure}
In Figs.~\ref{fig;se1} and \ref{fig;se2}, we plot the astronomical 
$S_{E1}$ and $S_{E2}$ factors for $E1$ and $E2$ transitions
of radiative $\alpha$ capture on $^{12}$C
as a function of the energy $E$ of $\alpha$-$^{12}$C 
in the center-of-mass frame by using the large
and small values of $h_R^{(1)}$ and $h_R^{(2)}$ with $r_C=0.01$~fm 
in Tables \ref{table;h1R} and \ref{table;h2R}, respectively.   
Experimental data of the $S$ factors are displayed in the figures as well. 
A minor cutoff dependence is observed in the plots. 
(Figures for the other $r_C$ values are suppressed.)
We note that the curves plotted in the figures are not fitted to the data of 
$S$ factors; those are parameter-free predictions. 
One can see that the curve of $S_{E1}$ factor 
with the large value of $h_R^{(1)}$ in Fig.~\ref{fig;se1} 
and that of $S_{E2}$ factor with the small value of $h_R^{(2)}$ 
in Fig.~\ref{fig;se2} relatively agree with the data.

In Fig.~\ref{fig;se1}, the data on the high-energy side 
are reproduced better, but those on the low-energy side are not.
%
%
It could be puzzling because a low-energy EFT is supposed to provide
a better description of a reaction at lower energy. 
This expectation of EFT stems from the decoupling theorem~\cite{dgh},
which describes how heavy particles at high energy enter into a low-energy
theory. In the present case, however, the discrepancy between the fitted 
curves and the data appears at the low energy side of the resonant peak.
Because a contribution from high energy will be more effective in the 
high energy side of resonant peak, to reproduce the shape of curves
on the low energy side of resonance, 
it would be necessary to take account of higher
order corrections to the $h_R^{(1)}$ term of $O\gamma O^*$ vertex 
function, which may play the role of form factor, a finite size effect of the 
nuclei involved in the reaction. 
The $S_{E1}$ factor at $E_G=0.3$~MeV is deduced as
$S_{E1} = 11.2~\textrm{keV}\,\textrm{b}$ with the large value of $h_R^{(1)}$
and 
$S_{E1} = 10.1~\textrm{keV}\,\textrm{b}$ with the small vale of $h_R^{(1)}$.
Those values are significantly smaller than the value 86.3 keV\,b recently 
estimated by means of an $R$ matrix analysis~\cite{detal-rmp17}.

In Fig.~\ref{fig;se2}, the sharp resonant peak at $E=2.68$~MeV 
is not reproduced by the curves because 
the propagator of resonant $2_2^+$ state of $^{16}$O
is not included.  
Moreover, the shape of the curve with a small value of $h_R^{(2)}$ 
at small energies, $E < 1.7$~MeV, is different from those one can find 
in literature; the curves of $S_{E2}$ previously reported continuously 
increase as the energy decreases; no maximum and minimum
points appear in the low-energy region. 
This is attributed to the fact that 
the effective range parameters for $l=2$ are not well determined
by using the phase shift data. 
It may be necessary to fit the effective range parameters by using 
the $S_{E2}$ data. 
While, the $S_{E2}$ factor at $E_G=0.3$~MeV is deduced as     
$S_{E2} = 1.3~\textrm{keV}\,\textrm{b}$ 
for the small value $h_R^{(2)}$
and $S_{E2} = 0.71~\textrm{keV}\,\textrm{b}$
for the large value of $h_R^{(2)}$.
Once again, those values are significantly smaller than 
the  value 45.3~keV\,b recently estimated by an $R$ matrix 
analysis~\cite{detal-rmp17}.

\vskip 2mm \noindent
{\bf 5. Results and discussion}

In the present work, we studied the radiative decay of sub-threshold 
$1_1^-$ and $2_1^+$ states of $^{16}$O within the cluster EFT.
We, first, discussed an improvement of the regularization method: 
we employed the dimensional regularization method to calculate 
the log-divergent term, which was regularized 
by using the cutoff regularization method in the previous work. 
Because the finite terms below the cutoff $r_C$ 
were ignored in the $r$-space integral, 
a minor cutoff dependence remains in the numerical result. 
In addition, we fixed the coupling constants, $y^{(0)}$, $y^{(1)}$, $y^{(2)}$,
of $O^*\alpha C$ and $\alpha CO$ vertex functions for the initial 
and final states of $^{16}$O by using the 
wave function normalization factors $\sqrt{{\cal Z}_l}$, 
equivalently the ANCs of $0_1^+$, $1_1^-$, $2_1^+$ 
states of $^{16}$O. The values of $\sqrt{{\cal Z}_l}$ (or those of the ANCs)
were deduced from the phase shift data of
elastic $\alpha$-$^{12}$C scattering at low energies. 
After fixing the coupling constants, $y^{(0)}$, $y^{(1)}$, $y^{(2)}$, only one
parameter, $h_R^{(1)}$ or $h_R^{(2)}$, remained in each of the radiative 
decay amplitudes, and we fitted them to the experimental data of 
decay rates of $1_1^-$ and $2_1^+$ states of $^{16}$O. 
We, then, perform a parameter-free calculation of $S$ factors for $E1$ and $E2$
transitions of $^{12}$C($\alpha$,$\gamma$)$^{16}$O 
by using the fitted parameters in the present study and found that the order
of the magnitude of $S$ factors, compared to the experimental data,
could be reproduced. 

For previous estimates of the $S$ factors in the literature, 
they are well summarized in Table IV in Ref.~\cite{detal-rmp17}.  
The reported values of $S_{E1}$ factors at $E_G=0.3$~MeV in the literatures 
are scattered from 3 to 340 keV\,b with various sizes of the error bars, 
and those of $S_{E2}$ factor at $E_G$ are from 5 to 220 keV\,b 
with the various error bars as well. 
As mentioned above, we have $S_{E1}=11.2$ and 10.1 keV\,b at $E_G$ with 
the large and small values of $h_R^{(1)}$ and $S_{E2}=1.3$ and 0.71 keV\,b with
the small and large values of $h_R^{(2)}$. Those values are found to be small,
an order of magnitude or two, compared to the large values of $S$ factors 
at $E_G$ reported in the literature. 
We note that those $S$ factors we obtained at 
$E_G$ are not fitted to the experimental values of $S$ factors in the 
energy range, $E=1$ to 3~MeV, displayed 
in Figs. \ref{fig;se1} and \ref{fig;se2}, 
in which one can see that the plotted curves do not reproduce
the experimental data well. 
To report a more reliable estimate of the $S$ factors at $E_G$ in cluster EFT,
it would be anticipated to reproduce the data of $S$ factors with a small
$\chi^2$ value. 

The ANCs (or the wave function normalization factors) can 
be a model-independent quantity when a clear separation scale exists.
For example, the asymptotic normalization factor of $s$-wave deuteron 
wave function works quite well~\cite{ah-prc05,ew-88} 
where the deuteron binding energy
is much smaller than the scale of $NN$ interactions, 
namely the pion mass, $m_\pi$. 
Among the three ANCs which we employed in the present work,
the ANC of $1_1^-$ state of $^{16}$O could be a model-independent one 
because the binding energy is so small, $B_1 = 0.045$~MeV, compared 
to the separation scale, $\Delta E = 4$~MeV. 
Our estimate of the ANC for the $1_1^-$ state of $^{16}$O turns out to 
be remarkably smaller, by about 20\% 
(about one sigma difference with experimental error bars),  
than those obtained from the 
$\alpha$-transfer reactions, ($^{6}$Li,$d$)~\cite{betal-prl99,aetal-prl15}.
Recently, a new deduction of the ANCs from the $\alpha$-transfer reactions 
is reported by considering the new estimate of ANC of $1_1^+$ state of
$^6$Li as a two-body $d$-$\alpha$ system~\cite{hetal-23}, 
where the ANC of $1_1^+$ state of $^{6}$Li is obtained
by means of an \textit{ab initio} calculation of $\alpha$($d$,$\gamma$)$^6$Li 
capture rate and elastic $d$-$\alpha$ scattering at energies below 
3~MeV~\cite{hetal-prl22}; the new value of ANC of $1_1^+$ state of $^6$Li
turns out to be significantly larger than the value previously employed.  
The reported new value of the ANC of $1_1^-$ state of $^{16}$O
is now reduced and in better agreement with the result presented 
in Eq.~(\ref{eq;ANC,11m}). 

Though the binding energy of $2_1^+$ state of $^{16}$O is also small,
$B_2=0.245$~MeV, there is a large uncertainty,
at least a factor of 2, to deduce the ANC of $2_1^+$ state of $^{16}$O,
mainly because of the tiny phase shifts at energies below the resonant
$2_2^+$ state of $^{16}$O. (The difficulty to fix the ANC of $2_1^+$ state
of $^{16}$O from the phase shift of elastic $\alpha$-$^{12}$C scattering
is discussed in Appendix in Ref.~\cite{a-prc23}.)
The ambiguity in the value of ANC could be compensated by the fitted 
value of $h_R^{(2)}$ in the radiative decay amplitudes.
However, as mentioned before,
the energy dependence of $S$ factor for $E2$ transition of 
$^{12}$C($\alpha$,$\gamma$)$^{16}$O is different from that of the experimental
data of $S$ factor in Fig.~\ref{fig;se2}. It may be necessary to fix 
the effective range parameters for $l=2$ by using the data of $S$ factor
of $E2$ transition
or those of the other reactions. 

The use of the ANC of ground $0_1^+$ state of $^{16}$O is questionable,
though it was encouraging to see that the value of ANC of $0_1^+$ state of
$^{16}$O obtained in Eq.~(\ref{eq;ANC,01p}) 
is close to that employed in the recent
$R$-matrix analysis~\cite{detal-rmp17} as well as those fitted to the 
$S$ factor of $E1$ transition of $^{12}$C($\alpha$,$\gamma$)$^{16}$O,
as discussed above.  
Because the length scale between $\alpha$ and $^{12}$C in the ground state is 
quite short, $\gamma_0^{-1}\simeq 1$~fm (and $\gamma_0$ is larger than
the large scale of the theory, $\Lambda_H$, $\gamma_0> \Lambda_H$),
it seems that 
a model dependence in the ANC of $0_1^+$ state of $^{16}$O is inevitable. 
Only the order of magnitude of the ANC may be reliable, and thus,
it would be practical to fit the ANC (or the coupling constant $y^{(0)}$) 
to the data of $S$ factors together with the coefficients of contact 
terms, $h_R^{(1)}$ and $h_R^{(2)}$.  

We found the double values of fitted parameters, $h_R^{(1)}$ and $h_R^{(2)}$
in Tables \ref{table;h1R} and \ref{table;h2R}. As mentioned, this implies 
a large cancelation between the contributions from loop diagrams
and counter terms. The size of the radiative decay amplitudes may be 
estimated by using a relation, 
$[h_R^{(l)}(L) - h_R^{(l)}(S)]/
[h_R^{(l)}(L) + h_R^{(l)}(S)]$, where $h_R^{(l)}(L/S)$ represent
the large/small values of $h_R^{(l)}$, and we have 0.025 for $l=1$
and 0.013 for $l=2$ by using the values at $r_C=0.01$~fm in the tables.  
Two orders of magnitude large values of $h_R^{(1)}$ and $h_R^{(2)}$ are
obtained, compared to the size of amplitudes for the radiative decay.
This could be a similar situation to that for the modification of counting 
rules introduced in the effective range parameters of 
elastic $\alpha$-$^{12}$C scattering at low energies. 
Thus, we might need to introduce higher-order corrections 
to the $h_R^{(1)}$ and $h_R^{(2)}$ contact vertex functions 
as counter terms. 
As discussed above,
such corrections would play a role of the form factor, which describe
a finite size effect of $^{16}$O.  

\vskip 2mm \noindent
{\bf Acknowledgements}

This work was supported by
the National Research Foundation of Korea (NRF) grant funded by the
Korean government (MSIT) (No. 2019R1F1A1040362 and 2022R1F1A1070060).

\vskip 2mm \noindent
{\bf Appendix A}

\begin{table}
\begin{center}
\begin{tabular}{ l | l }
\hline 
$m_{O^*(1_1^-)}$ & 14902.196426(140) MeV \cr
$ m_{O^*(2_1^+)}$ & 14901.9967(6)  MeV \cr
$m_{O}$ & 14895.079576 MeV \cr \hline
$m_\alpha$ & 3727.379 MeV \cr 
$m_C$ & 11174.862 MeV \cr \hline
\end{tabular}
\caption{
Masses of nuclei in the unit of MeV, which are used in the present work.  
}
\label{table;masses}
\end{center}
\end{table}
Masses of $^{16}$O*($1_1^-$), $^{16}$O*($2_1^+$), and $^{16}$O 
as well as $\alpha$ and $^{12}$C are
presented in Table \ref{table;masses},
and the mass differences (excited energies) of $1_1^-$ and $2_1^+$ 
states with respect to the ground $0_1^+$ state of $^{16}$O 
are~\cite{twc-npa93}
\bea
\Delta_1 &=& m_{O^*(1_1^-)} - m_O = 7.11685(14) \ \ \textrm{MeV}\,,
\\
\Delta_2 &=& m_{O^*(2_1^+)} - m_O = 6.9171(6) \ \ \textrm{MeV}\,.
\eea

At the rest frame for $^{16}$O$^*$, 
the $1_1^-$ and $2_1^+$ states of $^{16}$O decay 
to the ground state of $^{16}$O
by emitting a photon. 
The energy-momentum conservation including the terms 
up to $1/m_O$ corrections reads
\bea
\Delta_l &=& k_l + \frac{1}{2m_O}k_l^2 
\simeq k_l + \frac{1}{2m_O}\Delta_l^2\,,
\eea
with $l=1,2$. 
The magnitudes of photon momenta $k_1$ and $k_2$ 
(or photon energies) are obtained as 
\bea
k_1 &=& \Delta_1 - \frac{1}{2m_O}\Delta_1^2 = 7.1152(1)\ \ \textrm{MeV}\,,
\label{eq;k1}
\\
k_2 &=& \Delta_2 - \frac{1}{2m_O}\Delta_2^2 = 6.9155(6)\ \ \textrm{MeV}\,,
\label{eq;k2}
\eea
where $\Delta_1^2/(2m_O)\simeq 1.7\times 10^{-3}$~MeV and 
$\Delta_2^2/(2m_O)\simeq 1.6\times 10^{-3}$~MeV.

Thus, one has the spin-averaged decay rates as
\bea
\Gamma_l &=& 
\frac{k_l}{2\pi} 
\frac{1}{2S_{O^*(l)}+1}\sum_{spins}|A^{(l)}|^2\,,
\label{eq;Gamma_l}
\eea
where $S_{O^*(1_1^-)}=1$ and $S_{O*(2_1^+)}=2$, 
and $A^{(l)}$ are the radiative decay amplitudes which we calculate 
from the effective Lagrangian.

\vskip 2mm \noindent
{\bf Appendix B}

In the previous works, we calculated the ANCs of $0_1^+$ and $0_2^+$ 
states of $^{16}$O, not including the resonant states of $^{16}$O, 
by fitting the parameters to the phase shift of elastic $\alpha$-$^{12}$C 
scattering for $l=0$ below the energy of the first resonant $0_3^+$ 
state~\cite{a-jkps18}. 
We also calculated the ANC of $0_2^+$ state of $^{16}$O including 
the resonant $0_3^+$ and $0_4^+$ states of $^{16}$O (but not including the 
$0_1^+$ state of $^{16}$O) by fitting the parameters
to the phase shift data below the $p$-$^{15}$N breakup energy~\cite{a-prc23}. 
In this appendix, we discuss the derivation of the ANC of the ground $0_1^+$ 
state of $^{16}$O including the excited $0_2^+$ bound state and $0_3^+$
and $0_4^+$ resonant states of $^{16}$O by fitting the parameters to the 
phase shift below the $p$-$^{16}$N breakup energy.  

We employ the expression of $S$ matrix for elastic $\alpha$-$^{12}$C scattering
at low energies for $l=0$ in terms of the effective range parameters 
in Eq.~(28) in Ref.~\cite{a-prc23}: 
\bea
e^{2i\delta_0} &=& 
\frac{
K_0(p) -2\kappa \textrm{Re}H_0(p) + ipC_\eta^2
}{
K_0(p) -2\kappa \textrm{Re}H_0(p) - ipC_\eta^2
}
\prod_{i=3}^4
\frac{
E - E_{R(0i)} + R_{(0i)}(E) - i \frac12\Gamma_{(0i)}(E)
}{
E - E_{R(0i)} + R_{(0i)}(E) + i \frac12\Gamma_{(0i)}(E)
}\,,
\eea
where $\delta_0$ is the phase shift of elastic $\alpha$-$^{12}$C scattering
for $l=0$, and $K_0(p) = D_0(p) + 2\kappa H_0(p)$; $D_0(p)$ is given in 
Eq.~(\ref{eq;D0}) and has two zeros for the $0_1^+$ and $0_2^+$ bound states,
$D_0(i\gamma_{01})=0$ and $D_0(i\gamma_{02})=0$.  
Furthermore,
\bea
R_{(0i)}(E) &=& 
a_{(0i)}(E-E_{R(0i)})^2 
+ b_{(0i)}(E-E_{R(0i)})^3\,,
\\
\Gamma_{(0i)}(E) &=& 
\Gamma_{R(0i)}  \frac{pC_\eta^2}{p_rC_{\eta_r}^2}\,,
\ \ \ 
C_\eta^2 = \frac{2\pi\eta}{e^{2\pi\eta}-1}\,,
\eea
where $E_{R(0i)}$ and $\Gamma_{R(0i)}$ are resonant energies and widths,
and $a_{(0i)}$ and $b_{(0i)}$ are the coefficients of second order 
$(E-E_{R(0i)})^2$ and third order $(E-E_{R(0i)})^3$ contributions 
in the Breit-Wigner-like expression for the resonant states. 
$p_r$ are resonant momenta, $p_r = \sqrt{2\mu E_{R(0i)}}$ and 
$\eta_r =\kappa/p_r$. (We suppress the index $i$ for $p_r$ and $\eta_r$.)  

After including the two zeros in $D_0(p)$ in Eq.~(\ref{eq;D0}),
we have 10 parameters, 
$\{
P_0$, $Q_0;$ 
$E_{R(03)}$, $\Gamma_{R(03)}$, $a_{(03)}$, $b_{(03)};$
$E_{R(04)}$, $\Gamma_{R(04)}$, $a_{(04)}$, $b_{(04)}\}$,
in general, however, we exclude $a_{(03)}$ and $b_{(03)}$ 
from the fit because the resonant $0_3^+$ state is so sharp and the fit 
is insensitive to $a_{(03)}$ and $b_{(03)}$. 
We fix $E_{R(04)}$ and $\Gamma_{R(04)}$ by using the experimental data,
$E_{R(04)}^{exp} = 6.870(15)$~MeV and $\Gamma_{R(04)}^{exp}=185(35)$~keV
for the $0_4^+$ state~\cite{twc-npa93} 
because its peak does not appear in the data;
it provides a background contribution from high energy, where 
$a_{(04)}$ and $b_{(04)}$ describe the high energy tale from 
the $0_4^+$ state of $^{16}$O. 
Thus, we have
6 parameters, $\{P_0$, $Q_0$; $E_{R(03)}$, $\Gamma_{R(03)}$; 
$a_{(04)}$, $b_{(04)}\}$, to fit the phase shift data.      

The six parameters are fitted to the phase shift data of elastic 
$\alpha$-$^{12}$C scattering measured at $E_\alpha = 2.6$ -- $6.62$~MeV,
reported by Tischhauser et al.~\cite{tetal-prc09}, where $E_\alpha$ is 
the $\alpha$ energy in the lab frame, 
and we have
\bea
P_0 &=& -0.03452(2) \ \ \textrm{fm}^3\,,
\ \ \ 
Q_0 = 0.001723(7) \ \ \textrm{fm}^5\,,
\label{eq;P0Q0}
\\
E_{R(03)} &=& 4.8883(1) \ \ \textrm{MeV}\,,
\ \ \ 
\Gamma_{R(03)} = 1.35(3) \ \ \textrm{keV}\,,
\\
a_{(04)} &=& 0.756(7) \ \ \textrm{MeV}^{-1}\,,
\ \ \ 
b_{(04)} = 0.167(4) \ \ \textrm{MeV}^{-2}\,,
\eea
where the chi-square per the number of data is $\chi^2/N = 0.10$:
$N$ is the number of data, $N=252$. 
The fitted values of $E_{R(03)}$ and $\Gamma_{R(03)}$ agree well with 
the experimental values, $E_{R(03)}^{exp}=4.887(2)$~MeV and 
$\Gamma_{R(03)}^{exp}=1.5(5)$~keV~\cite{twc-npa93}. 

\begin{figure}
\begin{center}
  \includegraphics[width=13cm]{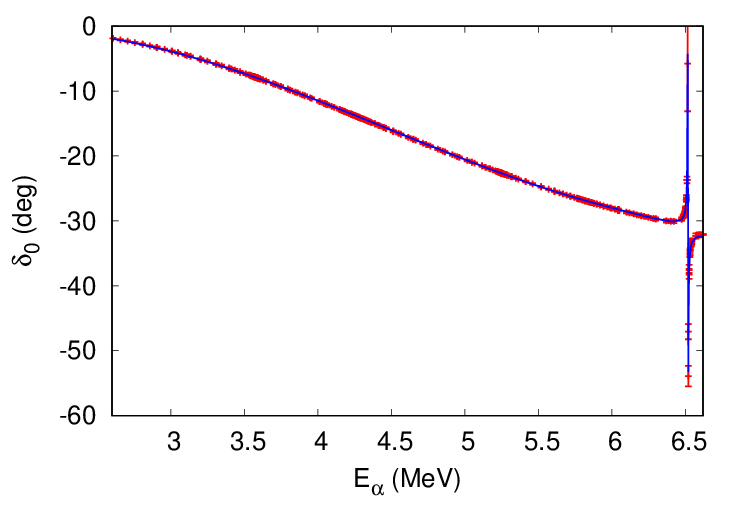}
\caption{
Phase shift $\delta_0$ of elastic $\alpha$-$^{12}$C scattering for 
$l=0$ is plotted as a function of $E_\alpha$ by using the fitted parameters,
where $E_\alpha$ is the $\alpha$ energy in the lab frame. 
The experimental data are displayed in the figure as well. 
}
\label{fig;del0}       
\end{center}
\end{figure}
In Fig.~\ref{fig;del0}, we plot the phase shift, $\delta_0$, 
using the fitted 
parameters and include the experimental data in the figure.
One can see that the plotted curve reproduces the data well. 
\begin{table}
\begin{center}
\begin{tabular}{ l | c c c  }
  & w/o $0_3^+$, $0_4^+$\cite{a-jkps18} & w/o $0_1^+$\cite{a-prc23} & 
 This work \cr \hline
$|C_b|_{01} (\textrm{fm}^{-1/2})$ & 44(2) & -- & 44.5(3) \cr
$|C_b|_{02} (\textrm{fm}^{-1/2})$ & 605(44) & 370(25) & 621(9) \cr \hline
\end{tabular}
\caption{
ANCs of $0_1^+$ and $0_2^+$ states of $^{16}$O, $|C_b|_{01}$ and 
$|C_b|_{02}$, respectively,  for two-body $\alpha$-$^{12}$C
bound state. Values in the second column are the result not including the 
resonant $0_3^+$ and $0_4^+$ state in the parameter fit~\cite{a-jkps18}.
A value in the third column is the result not including the ground $0_1^+$ 
state~\cite{a-prc23}.
Values in the last column are the result of this work. 
}
\label{table;Cbs}
\end{center}
\end{table}
In Table \ref{table;Cbs}, 
we display the values of ANCs of $0_1^+$ and $0_2^+$ states of $^{16}$O
for the two-body $\alpha$-$^{12}$C state, $|C_b|_{01}$ and $|C_b|_{02}$,
respectively, calculated by using the values of effective range parameters, 
$P_0$ and $Q_0$, in Eq.~(\ref{eq;P0Q0}). We also display 
the ANCs by using the parameters fitted not including the resonant $0_3^+$
and $0_4^+$ states and those not including the ground $0_1^+$ state in 
the table.  One can see that the results of this work and those not including
the resonant $0_3^+$ and $0_4^+$ states are in good agreement within the
error bars, while the result of ANC for $0_2^+$ state obtained not including
the ground $0_1^+$ state is small, about 60\% of that of this work.

\end{document}